\documentclass[twocolumn,showpacs,preprintnumbers,amsmath,amssymb,superscriptaddress,prb]{revtex4}
\usepackage{graphicx}% Include figure files
\usepackage{dcolumn}% Align table columns on decimal point
\usepackage{bm}% bold math
\begin{document}

%\preprint{APS/123-QED}

\title{Effective temperature for hopping transport in a Gaussian DOS}% Force line breaks with \\

\author {F. Jansson}
\email {fjansson@abo.fi}
\affiliation{Graduate School of Materials Research, \AA bo Akademi University, Finland}
\affiliation{Department of Physics and Center for Functional Materials, \AA bo Akademi University, Finland}
%\affiliation{Department of Physics and Material Sciences Center, Phillips-University Marburg, Germany}
\author {S. D. Baranovskii}
\affiliation{Department of Physics and Material Sciences Center, Phillips-University Marburg, Germany}
\author{F. Gebhard}
\affiliation{Department of Physics and Material Sciences Center, Phillips-University Marburg, Germany}
\author {R. \"Osterbacka}
\affiliation{Department of Physics and Center for Functional Materials, \AA bo Akademi University, Finland}

\date{\today}
             
\begin{abstract}
For hopping transport in disordered materials, the mobility of 
charge carriers is strongly dependent on temperature and the 
electric field. Our numerical study shows that both the energy 
distribution and the mobility of charge carriers in systems with 
a Gaussian density of states, such as organic disordered 
semiconductors, can be described by a single parameter -- 
effective temperature, dependent on the magnitude of the electric 
field. Furthermore, this effective temperature does not depend on 
the concentration of charge carriers, while the mobility does 
depend on the charge carrier concentration. The concept of the 
effective temperature is shown to be valid for systems with and 
without space-energy correlations in the distribution of 
localized states.
\end{abstract}

\pacs{72.20.Ht, 72.20.Ee, 72.80.Le, 72.80.Ng}
%\keywords{Hopping transport, Effective temperature}%Use showkeys class option if keyword
%72.20.Ht High-field and nonlinear effects
%72.20.Ee Mobility edges; hopping transport
%72.80.Le Polymers; organic compounds (including organic semiconductors)
%72.80.Ng Disordered solids

\maketitle

\section{Introduction}
\label {sec-introduction} 
For many years much attention of
researchers has been devoted to the effect of high electric fields
on the hopping mobility of charge carriers in organic disordered
materials, such as conjugated and molecularly doped polymers and
semiconducting organic glasses.\cite{Bassler1993,Pope1999,Hadziioannou2000,Baranovski2006} 
Two peculiarities in the dependence of the carrier mobility on the
electric field are usually discussed in the literature. One of
them is the apparent decrease of the mobility with rising field at
relatively low fields and high temperatures reported in several
time-of-flight studies.\cite{Peled1988,Schein1992,Abkowitz1992,Bassler1993} 
The other one is the very strong non-linear increase of the carrier mobility
with electric field at low temperatures 
and high fields.\cite{Bassler1993,Pope1999,Hadziioannou2000,Baranovski2006} 
While the former peculiarity has been attributed
\cite{Hirao1995,Baranovski2006} to misinterpretation of
experimental data, the latter one has been confirmed in numerous
experimental studies and it is currently in the focus of
intensive theoretical research. Before we turn to discussing this
research, it is worth noting that very similar non-linear effects
with respect to the applied electric field have been known since decades
for transport phenomena in the inorganic noncrystalline materials,
such as amorphous semiconductors. Indeed, strong nonlinearities in
amorphous semiconductors were observed for the field dependence of
the dark conductivity,\cite{Nebel1992,Nagy1993} of the
photoconductivity,\cite{Stachowitz1990} and of the charge carrier
drift mobility \cite{Nebel1992,Antoniadis1991,Murayama1992} at
high electric fields. While the field-dependent hopping
conductivity at low temperatures was always a challenge for
a theoretical description, the theories for temperature dependence
of the hopping conductivity at low electric fields were
successfully developed for all transport regimes listed above (see
for instance Ref.~\onlinecite{Baranovski2006} and references therein).
Furthermore, it has been shown that the effect of a strong
electric field on transport coefficients in amorphous
semiconductors can be effectively described by replacing the
temperature parameter in the formulas for the low-field
temperature-dependent mobility and conductivity by an effective
tempera\-ture $T_\text{eff} (T, F)$, dependent on the magnitude of
the electric field $F$. Shklovskii \cite{Shklovskii1973} was the
first who recognized that, for hopping conduction, a strong
electric field plays a role similar to that of temperature. 
In the presence of the field, the number of sites
available for charge transport is essentially enhanced in the
direction prescribed by the field. The distance between sites
available for hopping transport shortens and hence electrons can
move faster.\cite{Shklovskii1973} The concept of the
effective temperature has been studied in detail for systems with
an exponential energy distribution of localized states usually
assumed for inorganic noncrystalline materials:
\begin {equation}
\label {eq-exp-dos}
 g (\varepsilon) = \frac {N} {\sigma} \exp
\left(\frac {\varepsilon} {\sigma} \right),
\end {equation}
where $N$ is the concentration of localized states and $\sigma$ is
the energy scale of the distribution. By studying the steady-state
energy distribution of electrons in numerical calculations and
computer simulations \cite{Marianer-Shklovskii,Baranovskii1993}
and by computer simulations of the steady-state
hopping conductivity and the transient energy relaxation of
carriers \cite{Cleve1995} the following result has been found. The
whole set of transport coefficients can be represented as a
function of a single parameter $T_\text{eff} (T, F)$:
\begin {equation}
\label {eq-Teff} T_\text{eff} = \left[ T^\beta + \left(\gamma
\frac{e F a} {k} \right)^\beta \right]^{1/\beta}
\end {equation}
with $\beta=2$ and values of $\gamma$ in the range $0.5$ - $0.9$
depending on which transport coefficient is considered.\cite{Cleve1995} 
In this expression $a$ is the localization
length of charge carriers in the localized states, $e$ is the
elementary charge, and $k$ is the Boltzmann constant. Herewith
the problem of nonlinearities of transport coefficients with
respect to the applied electric field for inorganic noncrystalline
materials with the density of states (DOS) described by 
Eq.~(\ref{eq-exp-dos}) has been solved.

Let us now turn to the organic disordered materials. In such
systems the density of states involved in the hopping transport of
charge carriers is believed to be not purely exponential
as in Eq.~(\ref{eq-exp-dos}), but rather to be described by a Gaussian
energy distribution:
\cite{Bassler1993,Hartenstein1995,Schmechel2002,Pope1999,Hadziioannou2000,Baranovski2006}

\begin {equation}
\label {eq-gaus-dos} g (\varepsilon) = \frac {N} {\sigma \sqrt{2
\pi}} \exp \left( - \frac {\varepsilon^2} {2 \sigma^2} \right).
\end {equation}

Numerous computer simulations have been devoted to studying the
field nonlinearities of transport coefficients in the hopping
regime in such systems. Two models of disordered organic materials
were considered: the so-called Gaussian disorder model (GDM)
suggested by B\"assler et al.\cite{Bassler1993}\ and the so-called
correlated disorder model (CDM) considered by Garstein and Conwell,
\cite{Garstein1995} by Dunlap et al.,\cite{Dunlap1996} and by
Novikov et al.\cite{Novikov1998a,Novikov1998b} 
In both models, the field dependence of carrier mobility has been studied by 
computer simulations.
While analytical
calculations have been carried out in order to justify the CDM,
\cite{Dunlap1996,Novikov1998a,Novikov1998b} a consistent
analytical theory for the field dependence of the hopping mobility
in a Gaussian DOS is still missing.

It is tempting to try to apply the concept of the effective
temperature to organic disordered systems with the Gaussian DOS
described by Eq.~(\ref{eq-gaus-dos}), since this concept has
proven to be very successful for inorganic systems with
the exponential DOS described by Eq.~(\ref{eq-exp-dos}). The idea that
a strong electric field leads to heating of the charge carriers in
organic materials has already been considered in several
theoretical studies. Recently Preezant and Tessler \cite{Preezant2006} 
performed such a study in the framework of an analytical
approach that artificially decouples the energy-dependent
factors in the hopping transport from the space-dependent
factors. However, this approach has been shown
\cite{Baranovski2006,Coehoorn2005} to be unsuitable for
describing hopping transport processes in disordered
materials. Moreover, such an approach leads to an
effective temperature that essentially differs from the one
considered in previous studies for inorganic materials
\cite{Shklovskii1973,Marianer-Shklovskii,Baranovskii1993,Cleve1995,Baranovski2006}
where the effective temperature arises from the interplay between
spatial- and energy dependent factors in hopping processes. 
Li, Meller and Kosina\cite{Li2007} recently approached this problem 
by inserting the dependence of transition rates on the electric field 
and that of the percolation threshold into the percolation theory 
of Vissenberg and Matters \cite{Vissenberg1998}.
More recently, Limketkai et al.\cite{Limketkai2007} exploited the
concept of the effective temperature in order to account for the
strong field nonlinearity of the conductivity and carrier mobility
observed in organic semiconductors.\cite{Bruetting2001}
Limketkai et al.\ chose the expression for the
effective temperature $T_\text{eff} (T, F)$ in the form of Eq.~(\ref{eq-Teff}) 
with $\beta=1$ and $\gamma=0.5$. It has
been proven however\cite{Cleve1995} that such expression for the
effective temperature with $\beta=1$ cannot be considered as suitable. Indeed,
suppose the conductivity $G$ is dependent on $T_\text{eff} (T, F)$
solely. Then
\begin {equation}
\label {eq-dG} \frac{dG}{dF} = \frac {dG}
{dT_\text{eff}}\frac{dT_\text{eff}}{dF} .
\end {equation}
In the Ohmic transport regime at $F\ll kT/ea$, the conductivity
$G$ must be field independent, implying that
\begin {equation}
\label{eq-Fto0}
\frac {d T_\text{eff}}{d F} \rightarrow 0 \text{ as } F
\rightarrow 0.
\end {equation}
The function described by Eq.~(\ref{eq-Teff}) with $\beta=1$
obviously does not fulfill this condition. However, any function
of this kind with $\beta > 1$ is consistent with Eq.~(\ref{eq-Fto0}) 
along with the necessary condition $T_\text{eff} =
T$ at $F = 0$ and $T_\text{eff} \propto F$ at $T = 0$.

Therefore the challenging problem arises to find out whether the
concept of the effective temperature is applicable to systems with
a Gaussian DOS and if yes, what is the expression for
$T_\text{eff} (T, F)$. We try to answer these questions in the
present study. For this purpose we follow the idea of Marianer and
Shklov\-skii \cite {Marianer-Shklovskii} to look at the energy
distribution of charge carriers using the numerical method of
nonlinear balance equations suggested by Yu et al.\cite{Yu2000,Yu2001,Pasveer2005,Cottaar2006}
and apply it to electron transitions in a Gaussian DOS. Preliminary data for the 
energy distribution of charge carriers in the GDM at low carrier 
concentrations confirmed the concept of the effective temperature.\cite{Jansson2008}

The paper is organized as follows. In Sec.~\ref{sec-numerical} the
numerical method used in our study is described. In 
Sec.~\ref{sec-distribution} the results for the energy
distribution function of charge carriers in the GDM are presented. It is shown
that at finite temperatures and electric fields the energy
distribution of charge carriers in a Gaussian DOS can be well
described by the Fermi-Dirac distribution function characterized
by $T_\text{eff} (T, F)$ given by Eq.~(\ref{eq-Teff}) with
$\beta = 1.54 \pm 0.2$ and $\gamma = 0.64 \pm 0.2$
thus confirming
the applicability of the concept of the effective temperature to
organic disordered solids. Furthermore, we show that the
expression for $T_\text{eff} (T, F)$ is stable against changes in
the concentration of charge carriers. Numerical results obtained
for the mobility of charge carriers also confirm the concept of the
effective temperature for the GDM. In Sec.~\ref{sec-correlation}
the corresponding results for the CDM are presented. They prove
that the concept of the effective temperature is valid also for
correlated systems. 
Concluding remarks are gathered in Sec.~\ref{sec-conclusions}.

\section{Numerical method}
\label {sec-numerical} In order to check the validity of the
concept of the effective temperature, we solved a system of
non-linear balance equations with respect to the steady-state
occupation probabilities $p_{i}$ of sites and checked, following
the idea of Marianer and Shklovskii, \cite {Marianer-Shklovskii}
whether this distribution can be fitted by a Fermi--Dirac
distribution
\begin {equation}
  \label{eq-FD}
  p(\varepsilon) = \frac {1}{1 + e^ {(\varepsilon-\mu_\text{c})/kT_\text{eff}}},
\end {equation}
with some single parameter $T_\text{eff} (T, F)$. We also
calculated the carrier drift velocity along the field direction
and the corresponding mobility at different temperatures and
electric fields and checked whether the mobility can be described
as a function of a single parameter $T_\text{eff} (T, F)$ combined
from the temperature $T$ and the field strength $F$.

The balance equation for the occupation probability $p_{i}$ of a
site $i$ has the form \cite{Yu2000,Yu2001,Pasveer2005,Cottaar2006}
%rate out = rate in
\begin{equation}
\label{eq-balance}
 \sum_{j \ne i} p_i \Gamma_{ij} (1-p_j) =
 \sum_{j \ne i} p_j \Gamma_{ji} (1-p_i),
\end{equation}
where the rate of jumping from site $i$ to site $j$ is given by
the Miller--Abrahams formula,
\begin {equation}
\Gamma_{ij} = \nu_0 \, e^{-2\frac{\Delta R_{ij}}{a}} \left\{
\begin{array}{ll}
 e^{ - \frac{\Delta \varepsilon_{ij}}{kT}} \qquad & ,\ \Delta \varepsilon_{ij} > 0\\
 1  & ,\ \Delta \varepsilon_{ij} \leq 0
  \end{array}  \right. .
\label{millerabrahams}
\end {equation}
Here $\Delta \varepsilon_{ij}$ is the difference between energies of 
states $j$ and $i$,
$\Delta R_{ij}$ is the distance between these states, and $\nu_0$ is 
the attempt-to-escape frequency.
While in the initial work of Marianer and Shklovskii \cite
{Marianer-Shklovskii} the limit of a low concentration of charge carriers
was considered and the balance equations were linearized, we study
a system with finite number of charge carriers $n$. In order to
solve the system of the nonlinear balance equations we use the
iterative numerical procedure suggested by Yu et al.\
\cite{Yu2000,Yu2001,Pasveer2005,Cottaar2006} 
The balance equation
(\ref{eq-balance}) is rewritten in the form
\begin{equation}
\label{eq-iter}
 p_i = \frac {\displaystyle \sum_{j \ne i} \Gamma_{ji}p_j} {\displaystyle \sum_{j \ne i} \Gamma_{ij} -
\sum_{j \ne i} (\Gamma_{ij} - \Gamma_{ji})p_j},
\end{equation}
where the right hand side does not contain $p_i$. This expression
is used iteratively to find a solution to Eq.~(\ref{eq-balance}).
In each iteration step, all $p_i$ are updated. If $p_j$ has
already been calculated in the current step, this value 
is used, otherwise the result from the previous step is used.
This so called implicit iteration is necessary for obtaining
convergence. \cite {Yu2001} In the sums over $j$, only the most
important transitions are needed. We have considered jumps shorter
than a cut-off length $R_\text{cutoff}$, chosen so large that 
it does not affect the transport parameters.

The procedure is repeated
until the relative change of any one of the probabilities is smaller than
$10^{-10}$. We note that the solution seems to converge faster if
the sites are placed on a lattice than at random. When the
localization length is small, more iterations are needed, but the
cut-off length can be reduced.

If an electric field $F$ is applied in $z$ direction, the
difference in energy between sites $j$ and $i$ is given by
\begin {equation}
\label{eq-delta} \Delta \varepsilon_{ij} = \varepsilon_j - \varepsilon_i -
F e (z_j - z_i).
\end {equation}

When the probability distribution of the charge carriers over the
sites is known, the drift velocity of the carriers along the field
and their mobility are given by
\begin {equation}
v_z  = \sum_{i, j \ne i} p_i \Gamma_{ij} (1-p_j) (z_j - z_i)/n, \qquad \mu = \frac {v_z} {F}.
\end {equation}

We studied a system of $M$ sites distributed randomly within a
cube with a side length $L$. Each site has a random energy
$\varepsilon$, from a Gaussian distribution with the width $\sigma$.
The density of states is given by Eq.~(\ref{eq-gaus-dos}).
Periodic boundary conditions were applied in all directions. 
%The carrier is allowed to jump only to the closest copy of each site,
%which gives a cut-off in jumping length as $L/2$. 
The calculations
were performed in dimensionless units, where the width of the DOS
$\sigma$, the site concentration $N=M/L^{3}$, the Boltzmann constant $k$, 
elementary charge $e$, and attempt-to-jump frequency $\nu_0$ are equal to unity.

We used a system containing $M=8000$ sites inside a cube with side
length $L=20$. The localization length was varied between $0.2$ and
$1$, while the particle concentration was varied in the interval
between $c = 10^{-5}$ and $c = 0.1$. 

For each choice of $T$, $F$, concentration $c$, and localization
length $a$, a number of realizations of the system were generated
and the steady state occupation probabilities were determined. For
each realization, the carrier mobility and effective temperature
were calculated. The magnitudes of the effective temperaures obtained 
with different realizations of the system
were equal to each other within the accuracy of one percent.

The effective temperature was determined from the occupation probabilities $p_i$ 
with a linear fit of $\ln (1/p_i)$ as a function of $\varepsilon_i$,
\begin {equation}
\ln \left( \frac{1}{p_i} \right) = \frac{1}{kT_\text{eff}} \varepsilon_i - \frac{\mu_\text{c}}{kT_\text{eff}}.
\end {equation}
This approach works well when the variation in occupation probabilities 
for sites with similar energies is small.
When the variation is larger, we found it better to 
produce a histogram of $p(\varepsilon)$ and then fit to this histogram as above.
The histogram approach seems more correct, since it uses the average occupation probability 
at a given energy, in contrast to the direct fit that uses the average of $\ln (1/p)$.
All results for effective temperature presented below were 
obtained with the histogram method.
The parameters $\beta$ and $\gamma$ in Eq.~(\ref{eq-Teff}) were
determined by simultaneous fitting of all data points to the
surface given by this equation.

In the next Section we present the results of calculations for a
system without any correlation between site energies and their
spatial positions (GDM). In Section \ref{sec-correlation} we
present the results of calculations for the CDM -- a system, in
which the energies of neighboring sites are correlated.
In order to introduce correlations, we generate an
initial energy for each site with a Gaussian distribution. We then
calculate the real energy for each site by averaging the initial
energies of all sites inside the correlation radius
$R_\text{corr}$, in accord with the recepie suggested by Garstein and Conwell.
\cite{Garstein1995}
This averaging procedure has two effects. It will
introduce a correlation in energy between spatially close sites
and it will decrease the width of the energy distribution. 
In order to keep the width of the energy distribution independent of the 
correlation length, the site-energies were rescaled to provide the initial
energy width $\sigma$.
We discuss these effects of space--energy correlations in more detail
in Section \ref{sec-correlation}.

At very high electric fields, when the energy landscape is
strongly sloped along the field direction and the energy
differences between the successive states become negative as given
by Eq.~(\ref{eq-delta}), the carrier drift velocity saturates and
becomes field-independent.\cite{Bassler1993} We study the range
of electric fields, which are less than the field at which the
velocity saturates.

\section {Effective temperature for systems without space-energy correlations}
\label {sec-distribution}

\begin{figure}
\includegraphics{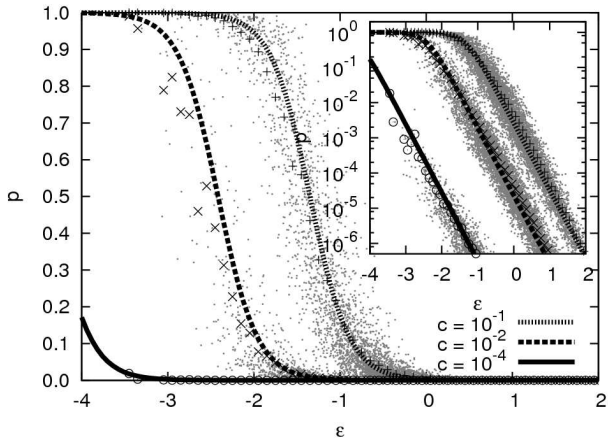}
\caption{Occupation probability as a function of site energy, for three different charge carrier concentrations $c$.
The temperature and field 
are $T = 0.2$ and $F = 0.4$, respectively. The dots show 
occupation probabilities of individual sites,
while the symbols show a histogram, i.e. average occupation 
probability for sites in an energy interval. The curves show Fermi--Dirac functions fitted to the histogram.}
\label {fig-occ}
\end{figure}

\begin{figure}
\includegraphics{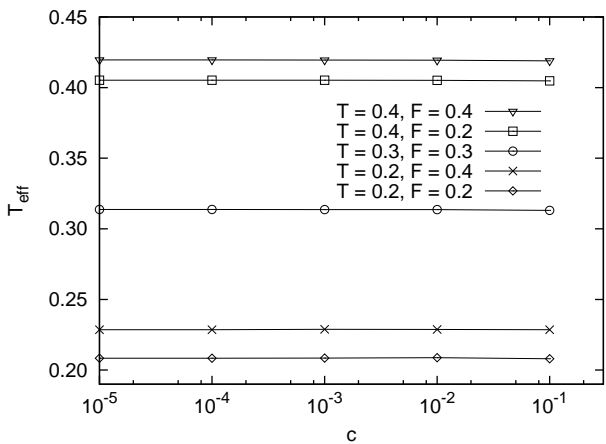}
\caption{Calculated $T_\text{eff}$ as a function of charge carrier concentration, for different values of $T$ and $F$.}
\label {fig-Teff-conc}
\end{figure}

In Fig.~\ref {fig-occ} the numerical results are shown for the 
occupation probability for each site, as a function of site energy,
for three different charge concentrations. The inset shows the same data 
on a logarithmic scale.

At low concentrations, the data can be fitted by a Boltzmann distribution 
with a parameter $T_\text{eff} (T, F)$, while the data 
at high concentrations show
a Fermi-Dirac shape, remarkably corresponding to the same
parameter $T_\text{eff} (T, F)$. 
Furthermore, Fig.~\ref{fig-Teff-conc} shows that the value of 
$T_\text{eff}(T,F)$ does not depend on the carrier concentration $c$.
These results are really remarkable because they
mean that the concept of the effective temperature is valid
for systems with a Gaussian density of states (so far it has been
proven only for exponential DOS described by 
Eq.~(\ref{eq-exp-dos})) and that the parameter
$T_\text{eff} (T, F)$ is universal with respect to the
concentration of charge carriers in the system under study. 

\begin{figure}
\includegraphics{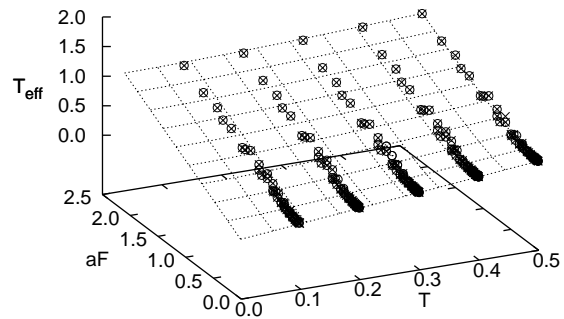}
\caption{$T_\text{eff}$ as a function of temperature $T$ and $aF$. 
Concentrations $c = 10^{-2}$ and  $10^{-5}$ are shown. The surface is the 
best fit to Eq.~(\ref{eq-Teff}), with $\beta = 1.54$ and $\gamma = 0.64$ }
\label {fig-Teff}
\end{figure}

In Fig.~\ref{fig-Teff} we bring together all the 
calculation results obtained for
various field strengths in the range $0 < F < 3$ and for various
temperatures in the range $0.1 < T <0.5$ for concentration of
carriers $c = 10^{-2}$ and  $10^{-5}$, in the form of a three-dimensional plot
of $T_\text{eff} (T, F)$, obtained by the
best fit of the calculated energy distributions by a Fermi-Dirac
function. 
Also the surface determined by Eq.~(\ref{eq-Teff}) with
parameters $\beta = 1.54; \gamma = 0.64$  is shown in this figure.
One can see an excellent agreement between the calculated results
and Eq.~(\ref{eq-Teff}) with these parameters. 

\begin{figure}
\includegraphics{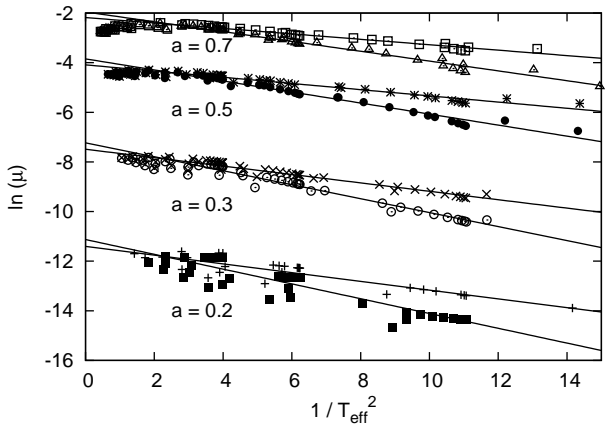}
\caption{The dependence of carrier mobility on the effective temperature, 
for different localization lengths $a$. In each pair of curves, the upper 
one is for charge carrier concentration $c = 10^{-2}$ and 
the lower one for $c = 10^{-5}$.}
\label {fig-mu-Teff}
\end{figure}

As described in the previous Section, we also studied the mobility
of charge carriers with respect to the applicability of the
concept of the effective temperature. In Fig.~\ref{fig-mu-Teff}
the dependences of
the carrier mobility as a function of $T_\text{eff} (T, F)$
determined from the fits of the energy distribution function are
shown for different concentrations $c$ of charge carriers. 
These data were obtained for various field strengths in the range $0 <
F < 3$ and for various temperatures in the range $0.1 < T < 0.5$.
The data were averaged over three realizations. For the chosen parameters, 
it appears not important 
whether the mobilities or the inverse mobilities were averaged.
While the magnitude of the mobility appears sensitive to the
concentration of carriers, in agreement with the results obtained
previously by several research groups, 
\cite{Yu2000,Yu2001,Cottaar2006,Pasveer2005,Pasveer2005b,Coehoorn2005,Vissenberg1998,Martens2003,Roichman2004,Baranovskii2002,Baranovskii2002b}
the magnitude of
the effective temperature for the given pair ($T, F$) does not
indicate any dependence on $c$.

Therefore one can conclude that for the GDM, i.e., for a
disordered system with a Gaussian distribution of energies and
without any correlations between spatial positions of hopping
sites and their energies, the concept of the effective temperature
is valid. In the next Section we present results for the CDM, i.e.\
for the correlated system.

\section {Effective temperature for systems with space-energy correlations}
\label {sec-correlation}

In this Section we study the effect of the space-energy
correlations on the interplay between the temperature $T$ and the
electric field $F$ with respect to the validity of the concept of
the effective temperature. 

\begin{figure}
\includegraphics{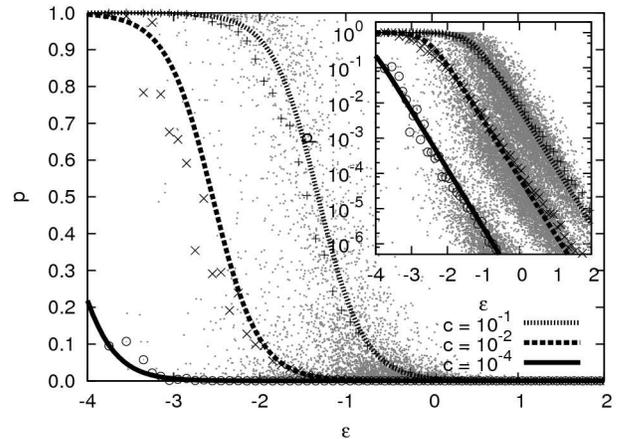}
\caption{Occupation probability in the correlated disorder case, 
with $R_\text{corr} = 2$.
The temperature and field 
are $T = 0.2$ and $F = 0.4$, respectively. The symbols are defined as in Fig.~\ref{fig-occ}.}
\label {fig-occ-corr}
\end{figure}

In Fig.~\ref {fig-occ-corr} the numerical results are shown for the 
occupation probability for each site as a function of site energy
for three different charge concentrations at correlation length $R_\text{corr} = 2$.
The inset shows the same data 
on a logarithmic scale.
In analogy to the uncorrelated system,
at low concentrations the data can be fitted by a Boltzmann distribution with a
parameter $T_\text{eff} (T, F)$, while the data at high concentrations show
a Fermi-Dirac shape corresponding to the same
parameter $T_\text{eff} (T, F)$ as for low concentration.

\begin{figure}
\includegraphics{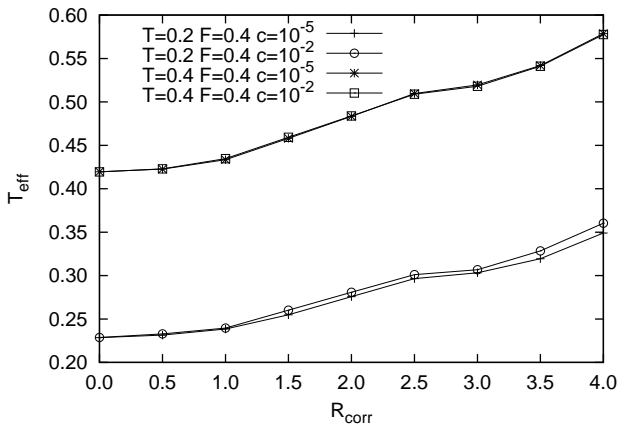}
\caption{The effective temperature as a function of the correlation length, for randomly placed sites, 
with $a = 0.5$.
}
\label {fig-Teff-Rcorr}
\end{figure}

Fig.~\ref {fig-Teff-Rcorr} shows the effective temperature as a function 
of the correlation length $R_\text{corr}$.
The effective temperature strongly depends on the correlation length, thus Eq.~(\ref {eq-Teff})
is not directly applicable to a system with space-energy correlations.

Before discussing the carrier mobility in a correlated system one should note the 
following. When the space-energy correlations are introduced as described in 
Sec.~\ref{sec-numerical}, the effective width $\sigma$ of the energy distribution 
of localized states decreases. Furthermore this width becomes smaller with 
increasing $R_\text{corr}$. Therefore one can expect 
that with increasing  $R_\text{corr}$ the mobility would increase just because of 
diminishing the energetic disorder due to the effect of correlations. In order 
to compare the carrier mobilities in systems with different correlations 
and the same energy disorder, one should therefore rescale the width of the energy 
distribution in the correlated systems and bring it to the value initially 
ascribed to the uncorrelated system.\cite{Garstein1995} The effect of 
correlations on the carrier mobility has already been studied by computer 
simulations, though in the frame of the lattice model and not with respect 
to the validity of the effective temperature.\cite{Garstein1995} In order 
to compare our results with those of previous studies we also have simulated 
mobilities in correlated systems within a lattice model along with simulations
in the random model.

\begin{figure}
\includegraphics{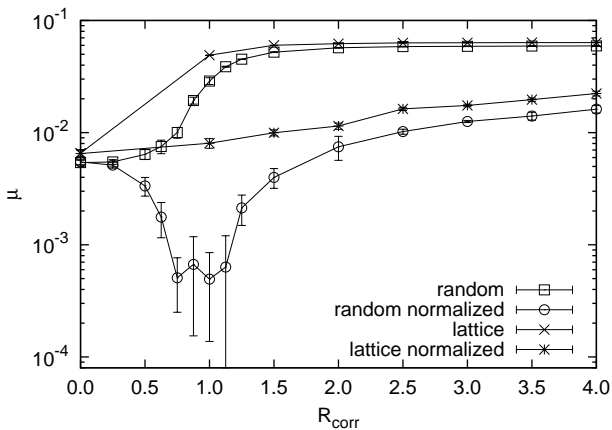}
\caption{Mobility as a function of correlation length, with and without rescaling of site energies, for
$T = 0.4$, $F = 0.4$, $a = 0.5$, $L = 30$.}
\label {fig-mu-Rcorr}
\end{figure}

\begin{figure}
\includegraphics{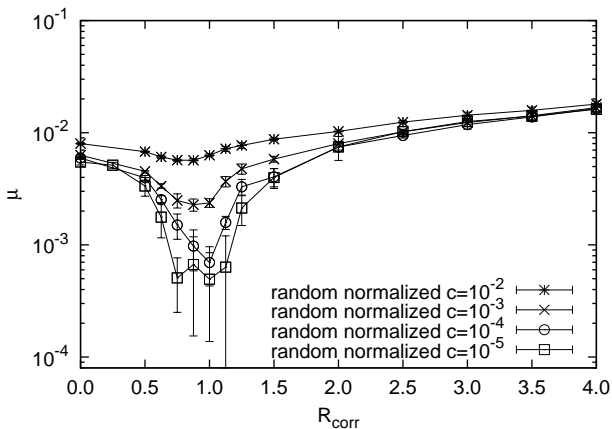}
\caption{Mobility as a function of correlation length, with randomly 
placed sites and rescaling of the site energies, 
for different concentrations of charge carriers;
$T = 0.4$, $F = 0.4$, $a = 0.5$, $L = 30$.}
\label {fig-mu-Rcorr-conc}
\end{figure}

Fig.~\ref {fig-mu-Rcorr} shows the mobility $\mu$ as a function
of the correlation length $R_\text{corr}$ with randomly placed sites and with sites on a lattice
for both cases: with and without rescaling (normalization) of the width of the energy distribution.
As expected, without rescaling the mobility always increases with the increase of the correlation 
length $R_\text{corr}$. However in the rescaled system there is an apparent difference between the 
lattice model and the random model with respect to the dependence 
of the mobility on the correlation length. While in the lattice model the mobility monotonously increases 
with $R_\text{corr}$ in accord with the results of  prevous studies,\cite{Garstein1995} in the random 
model the dependence of $\mu$ on $R_\text{corr}$ appears non-monotonous. At $R_\text{corr} \leq 1$
mobility decreases with $R_\text{corr}$. 
The reason for the low mobility at $R_\text{corr} \approx 1$ could be the following. 
The normalization (rescaling) procedure creates
a small number of sites with very low energies. While the general shape of the DOS for the correlated 
system is Gaussian, and the variance of the DOS is normalized to be equal to the variance of the uncorrelated system,
the DOS of the correlated system has longer tails, and the low-energy tail greatly reduces the carrier mobility. 
The sites in the tail are those sites that 
initially had a low energy, and have no neighbors within the distance $R_\text{corr}$. Thus they are 
unaffected by the averaging but are still rescaled with the normalization factor, which is 
typically close to 2 for $R_\text{corr} = 1$.
This idea is supported by data shown in Fig.~\ref{fig-mu-Rcorr-conc}. 
The dip in the mobility disappears when the charge carrier concentration is increased,
and the deep states are filled.

\section {Conclusions}
\label {sec-conclusions}
The concept of the effective temperature has been shown to be
valid for a system with Gaussian DOS with and without space-energy correlations in the 
distribution of localized states. From the numerical results for uncorrelated systems
one can conclude that the effective temperature is described by
Eq.~(\ref{eq-Teff}) with parameters  $\beta = 1.54 \pm 0.2 ; \gamma = 0.64 \pm 0.2$. Remarkably the 
validity of the concept of the effective temperature with very close numerical parameters has 
recently been reported on the basis of experimental study on the inorganic disordered 
material a-Si:H.\cite{Aoki2008} 
This quantitative agreement between experimental data and the results of our calculations
could mean that in the material studied in \cite{Aoki2008} the distribution of localized states can
be described by the Gaussian disorder model with uncorrelated distribution of localized states. 
Note that the effective temperature in our numerical calculations has been proven to be independent 
of the charge carrier concentration.

\begin{acknowledgments}
F.~J.\ wants to thank Prof.~P.~Bobbert for valuable discussion regarding the iterative solution method.
The calculations were performed at CSC, the Finnish IT center for science.
Financial support from the Academy of Finland project 107684 and
the TEKES NAMU project, from the Deutsche Forschungsgemeinschaft and
that of the Fonds der Che\-mischen Industrie is gratefully
acknowledged.
\end{acknowledgments}

\newpage

%\bibliography{teff-paper}% Produces the bibliography via BibTeX.

\end{document}